\documentclass[onecolumn,superscriptaddress,
amsmath,amssymb, aps, pre,]{revtex4-1}
\usepackage{dcolumn}
\usepackage{bm}
\usepackage{graphicx}
\usepackage{comment}
\usepackage[caption=false]{subfig}
\usepackage{color}

\newcommand{\bea}{\begin{eqnarray}}

\newcommand{\eea}{\end{eqnarray}}

\begin{document}

\begin{abstract}

We study the emergence of correlations between $N$ components of the position of a diffusive walker in $N$ dimensions 
that starts at the origin and resets to previously visited sites with certain probabilities. This is equivalent to $N$ 
independent one-dimensional diffusive processes starting from the origin and being subject to simultaneous resetting to positions 
visited in the past. Resetting follows a memory kernel that interpolates between resetting to the origin only, and the 
preferential relocation model, a path-dependent process which is highly non-Markov. For weak memory, the correlation 
coefficient between two components of the $N$-dimensional process grows monotonously with time and 
tends at late times to a constant bounded by $1/5$, the value corresponding to the non-equilibrium steady state of 
resetting to the origin. When memory is sufficiently long-ranged, the correlation is non-monotonous and reaches a 
maximum at a finite time before converging to its asymptotic value. These two regimes are separated by a critical 
memory parameter value. In the limiting case of the preferential relocation model, the components become uncorrelated 
at both short and long times, but the correlation vanishes  logarithmically slowly at late times. The emergence of 
correlations through resetting can be described in a unified way in all cases by noticing that the processes are 
conditionally independent and identically distributed, even in the presence of memory. In the non-Markovian case, the conditioning parameter is the duration of a Brownian path composed of several parts of the full 
trajectory of a fixed duration $t$.

\end{abstract}

\preprint{}
\title{Emerging correlations between diffusing particles evolving via simultaneous resetting with memory}
\author{Denis Boyer}
\email{Corresponding author, boyer@fisica.unam.mx}
\affiliation{Instituto de F\'\i sica, Universidad Nacional Aut\'onoma de M\'exico,
Ciudad de M\'exico 04510, M\'exico}
\author{Satya N. Majumdar}
\affiliation{LPTMS, CNRS, Univ.  Paris-Sud,  Universit\'e Paris-Saclay,  91405 Orsay,  France}

\maketitle

\section{Introduction}

Resetting is able to substantially alter the dynamics of stochastic processes through the emergence of non-equilibrium 
steady states (NESSs) and anomalous relaxation \cite{EM2011,P2015,MSS2015}. In the well-studied 
example of simple diffusion, the probability density function (PDF) of the position of a Brownian walker, that is 
interrupted with constant rate and instantaneously restarted from the initial position, tends at late times to a 
stationary distribution with a broken detailed balance and exhibiting exponential tails. Over the past decade, the 
characterization of NESSs in diffusive systems subject to resetting has been extended to a variety of resetting 
protocols \cite{EM2011b,NG2016,EM2016,PKR2019,BS2020,EMS2020}. Going beyond theoretical studies, diffusion with
stochastic resetting has also been realized experimentally in colloidal systems confined by optical 
traps~\cite{TPSRR2020,BBPMC2020,FBPCM2021}.

While the one-dimensional geometry has attracted much attention due to its relative simplicity, a body of research has also 
focused on the effects of resetting on diffusive processes in higher dimensions \cite{EM2011,EM2014,TPSRR2020,FBPCM2021}, or 
more generally on interacting systems with many degrees of freedom -- this includes fluctuating
interfaces~\cite{GMS2014}, two interacting Brownian particles~\cite{FE2017}, 
symmetric exclusion processes~\cite{BKP2019}, the Ising model evolving under Glauber dynamics~\cite{MMS2020}, the
Kuramoto model under subsystem resetting~\cite{MCG2024}, or ballistically moving
particles subject to threshold resetting~\cite{BMP2025} (see \cite{EMS2020} and \cite{NG2023} for reviews).
Only recently a quite remarkable feature of resetting in dimension larger than one has been unveiled~\cite{BLMS2023}.
Consider a single Brownian particle in $N$ dimensions with coordinates
$\vec x(t)= \{x_1(t),x_2(t),\ldots, x_N(t)\}$ that is stochastically reset to the origin with rate $r$. This problem is equivalent to $N$ independent one-dimensional Brownian motions that are  {\em simultaneously} reset with rate $r$ to their original values.
%All $N$ componentsof the position get {\em simultaneously} reset to their original values with rate $r$. 
Even though during the evolution between two
consecutive resetting events the processes $\{x_i(t)\}_{i=1,\ldots,N}$ evolve independently of each other, the simultaneous resetting 
introduces strong attractive and all-to-all correlations between them . There correlations grow with time before eventually
reaching a non-zero value in the stationary state~\cite{BLMS2023}. This phenomenon of ``dynamically induced correlation via 
simultaneous resetting" between independent particles has subsequently been found 
in a number of other classical models such as 
L\'evy flights and run-and-tumble particles undergoing simultaneous resetting~\cite{BLMS2024},
independent Ornstein-Uhlenbeck particles with a stochastically switching stiffness~\cite{BKMS2024} or a stochastically switching trap centre~\cite{SM2024}, independent Brownian particles subject to a common diffusing diffusivity~\cite{MMS2025}, or
Brownian particles subject to simultaneous resetting but with a non-Poissonian resetting protocol~\cite{DBMS2025}.
This mechanism for the emergence of strong correlations has recently been demonstrated experimentally in a system of $N$ colloidal particles
in a harmonic optical trap whose stiffness switches stochastically between two values~\cite{BCKMPS2025}.
Similar emergent correlations between independent degrees of freedom induced by simultaneous resetting
has also been found in quantum resetting models subject to quantum resetting to the initial state after a random exponentially distributed time. Examples include non-interacting spin chains in a transverse magnetic 
field~\cite{MCPL2022,SLP2025}, and noninteracting bosonic oscillators evolving unitarily after a sudden
quench of either the center of the trap~\cite{KMS2024} or the stiffness of the trap~\cite{MKMS2025}.

Processes under simultaneous resetting belong to a wider class of problems with a non-factorizable joint distribution in the 
stationary state, where the random variables are conditionally independent and identically distributed (c.i.i.d.). In the 
context of resetting to a fixed position in space (e.g. to the initial position), conditioned on the 
time interval since the last resetting event, the $N$ variables are actually 
independent. However, taking the average over this time elapsed since the last resetting event 
generates attractive all-to-all correlations between 
pairs. The c.i.i.d. structure is advantageous, as it allows the exact calculation of many physical observables which are 
usually not easy to compute in strongly correlated systems, such as the average density of particles, the distribution of the 
extremes, the consecutive spacings between particles, or the full counting statistics~\cite{BLMS2023,BLMS2024}.

Most of these recent studies on dynamically emergent correlations have considered memory-less stochastic resetting, which 
occurs to a fixed position or state, and does not depend on the history of the process. The aim of the present study is to 
investigate the emergence of correlations in a non-Markovian system where resetting occurs more generally to positions 
visited in the past. Consider for instance an animal in two dimensions with coordinates $\{x_1(t),x_2(t)\}$ that diffuses 
freely in its environment and revisits from time to time familiar places, i.e., positions visited at previous times. A 
modelling framework for this type of systems is the preferential relocation model or 'monkey walk', where any past position 
can be revisited via resetting, with a probability proportional to the total amount of time spent at that position by the 
walker since the initial time \cite{BS2014,BEM2017,MU2019}. Contrary to the case of resetting to the origin only, this 
process does not exhibit a NESS at late times but a very slow, logarithmic diffusion. The model also reproduces 
quantitatively several movement statistics of non-human primates observed in the wild \cite{BS2014}. While these memory 
processes have been studied theoretically mostly in one dimension, the existence of correlations induced dynamically 
between the two coordinates $x_1(t)$ and $x_2(t)$ is of possible interest in Ecology and animal foraging. The use of memory by 
animals in movement has been documented in many species and can be characterised, for instance, from the analysis of long-term trajectories obtained from tracking devices
equipping individuals reintroduced in a new environment \cite{GM2005,SKWB2010,F2013,MFM2014,FBMFM2021,R2021,RCM2022}. 
The study of correlations thus represents a possible way of identifying the use of memory and preferential relocation processes. The
goal of this paper is to compute exactly how the correlations between different components evolve with time
in the `monkey walk' model in $N$ dimensions. One of the main results is a striking
non-monotonic behaviour of these mutual correlations. The correlations between different components initially
grow with time and then eventually decreases to zero very slowly at large times. This feature is very different from the dynamics of mutual correlations in systems with resetting to a single position, where the
correlations grow with time monotonically and eventually saturates to a constant value~\cite{DBMS2025}.

The rest of the article is organized as follows. In Section \ref{sec:corrgen}, we introduce a correlation
coefficient that quantifies 
mutual correlations between two processes and show how it can be obtained from the small $k$ expansion of the Fourier 
transform of their joint PDF. In Section \ref{sec:model} we apply this method to compute
the mutual correlations in a resetting model 
with memory in $N$ dimensions, that is characterized by an exponential memory 
kernel with parameter $\lambda$. Section 
\ref{sec:lambdainfty} presents results on the limiting case of resetting to the origin ($\lambda\to\infty$), while Section 
\ref{sec:lambda0} is devoted to the analysis of the opposite limit $\lambda\to 0$ corresponding to the preferential 
relocation model. In Section \ref{sec:anylambda}, we compute the correlations in the general case at $\lambda$ finite, first 
in the infinite time limit (\ref{sec:lamb_infinite_t}) and then at finite times (\ref{sec:lamb_finite_t}). In Section 
\ref{sec:renewallike}, we show that models with resetting and memory generally admit a hidden c.i.i.d. structure, that 
extends the standard renewal approach of memory-less resetting processes and allows us to describe the emergence of 
correlations in a unified way in these systems. Conclusions are presented in Section \ref{sec:concl} and technical details 
in Appendix \ref{sec:appA}.

\section{Quantifying correlations between two processes}\label{sec:corrgen}

Let us consider a process $\vec{x}(t)$ in $N$ dimensions with components $\{x_i(t)\}_{i=1,...,N}$ 
that have the same initial condition $x_i(t=0)=0$ and 
the same marginal PDF $p(x_i,t)$. We focus on the correlations between two  components, say $x_1(t)$ and $x_2(t)$. The natural measure of correlations among two variables
is the $2\times 2$ matrix defined by 
\begin{equation}
C_{ij}(t) = \langle x_i(t) x_j(t)\rangle - \langle x_i(t)\rangle \langle x_j(t)\rangle\, ,\quad i=\{1,2\}\,, \ j=\{1,2\}\,.
\label{two_point_corr.1}
\end{equation}
We first notice that if the joint PDF $P(\vec x, t)$ is
invariant under permutation of the labels of the axis or particles, for instance due to isotropy, the correlation $C_{ii}(t)$ is the same for all $i$, while
$C_{ij}(t)$ is also the same for any pair of indices $(i,j)$ with $i\neq j$.
Secondly, the diffusive problems considered in the next Sections are also symmetrical under an inversion $x\to -x$, hence the mean
$\langle x_i(t)\rangle =0$ at all times $t$ for any $i$. In addition, the function 
$C_{ij}(t)$ with $i\neq j$ is also identically zero by symmetry at all times $t$, as the points $(x_1,x_2)$ and $(-x_1,x_2)$ turn out to have the same probability density. Hence, to detect a non-zero correlation between $x_1(t)$ and $x_2(t)$, it is necessary to focus on another $2\times2$ correlation matrix, such as
\begin{equation}
A_{ij}(t)= \langle x_i^2(t) x_j^2(t)\rangle - \langle x_i^2(t)\rangle\, \langle x_j^2(t)\rangle\, ,\quad i=\{1,2\}\,, \ j=\{1,2\}\,. 
\label{corr_2.1}
\end{equation}
Let us call $\lambda_+(t)$ and $\lambda_-(t)\le\lambda_+(t)$ the  eigenvalues of ${\mathbf A}$, and define the time-dependent correlation coefficient of the two processes as
\begin{equation}
a(t)\equiv\frac{\lambda_+(t)-\lambda_-(t)}{\lambda_+(t)+\lambda_-(t)}\, .
\label{corrcoeff1}
\end{equation}
Since $x_1(t)$ and $x_2(t)$ are identically distributed, $A_{11}(t)=A_{22}(t)=\langle x_1^4(t)\rangle-\langle x_1^2(t)\rangle^2$ and $A_{12}(t)=A_{21}(t)=\langle x_1^2(t)x_2^2(t)\rangle-\langle x_1^2(t)\rangle^2$. After straightforward algebra, the parameter $a(t)$ reads
\begin{equation}
a(t)=\frac{\langle x_1^2(t)x_2^2(t)\rangle-\langle x_1^2(t)\rangle^2}{\langle x_1^4(t)\rangle-\langle x_1^2(t)\rangle^2}\, .
\label{corrcoeff2}
\end{equation}
This quantity lies in the interval $[0,1]$, with the limiting cases $a(t)=0$ if the processes are independent at all times, while $a(t)=1$ for perfectly correlated processes, i.e., $x_2(t)=x_1(t)$ or $x_2(t)=-x_1(t)$.

In the examples to be analysed below, the Fourier transform of the joint distribution,
\begin{equation}
\tilde{P}(\vec k, t)=\int d\vec{x}\, e^{i\vec{k}\cdot\vec{x}}P(\vec{x},t)\, ,
\label{FT.1}
\end{equation}
only depends on the components $\{k_i\}$ of $\vec{k}$ through its square modulus $k^2= k_1^2+k_2^2+\ldots+ k_N^2$ at all times $t$. Let us suppose that $\tilde{P}(\vec k, t)$ can be expanded in power series at small $k^2$, 
\begin{equation}
\tilde{P}(\vec k, t)=1-c_2(t)\, k^2+c_4(t)\,k^4+\ldots\, ,
\label{Pkexpand}
\end{equation}
where the term of order $k^0$ comes from the normalization of $P(\vec{x},t)$. Clearly, $C_{12}(t)$ in Eq. (\ref{two_point_corr.1}) vanishes identically at all times $t$. This follows from the fact that
\begin{equation}
\langle x_1(t)x_2(t)\rangle = -\left.\frac{\partial^2 \tilde{P}(\vec k,t)  }{\partial k_1\, \partial k_2}
\right|_{\vec k=\vec 0}=0\, , 
\label{small_k.1}
\end{equation}
where the last equality is a consequence of Eq. (\ref{Pkexpand}). We similarly deduce the general relations,
\begin{eqnarray}
&&\langle x_1^2(t)\rangle=-\left.\frac{\partial^2 \tilde{P}(\vec k,t)  }{\partial k_1^2}
\right|_{\vec k=\vec 0}=2\,c_2(t)\, ,  \label{x2}\\
&&\langle x_1^4(t)\rangle=\left.\frac{\partial^4 \tilde{P}(\vec k,t)  }{\partial k_1^4}\right|_{\vec k=\vec 0}=24\,c_4(t)\, , 
\label{x4}\\
&&\langle x_1^2(t)x_2^2(t)\rangle= \left.\frac{\partial^4 \tilde{P}(\vec k,t)  }{\partial k_1^2\, \partial k_2^2}\right|_{\vec k=\vec 0}=8\,c_4(t)\, , \label{x2y2}
\end{eqnarray}
and obtain from Eq. (\ref{corrcoeff2}), 
\begin{equation}
a(t)=\frac{2c_4(t)-c_2^2(t)}{6c_4(t)-c_2^2(t)}\, .
\label{ac2c4}
\end{equation}
Hence the knowledge of $c_2(t)$ and $c_4(t)$ yields the correlation coefficient $a(t)$, which is the main quantity of interest here. Note that due to the isotropic structure of Eq. (\ref{Pkexpand}) and the subsequent relations (\ref{x4})-(\ref{x2y2}), one has $\langle x_1^2(t)x_2^2(t)\rangle=\frac{1}{3}\langle x_1^4(t)\rangle< \langle x_1^4(t)\rangle$. Therefore the variables $x_1$ and $x_2$ cannot be perfectly correlated and $a(t)<1$. 

\section{Diffusion with simultaneous resetting and a memory kernel $\phi(\tau)=e^{-\lambda \tau}$}\label{sec:model}

We consider a particle diffusing in $N$ spatial dimensions and starting at the origin [or, equivalently, $N$ independent particles diffusing on a line with initial positions $x_i(t=0)=0$, with $i=1,\ldots, N$]. With rate $r$, the particle resets to a previously visited position according
to the following protocol. The particle keeps track of its own trajectory $\vec{x}(\tau)$
for $0\le \tau\le t$. In a small time interval $[t,t+\Delta t]$, with probability $r\Delta t$ the particle chooses a previous time instant $\tau$ distributed with a PDF $K_t(\tau)$ and resets to the position occupied at that time $\tau$, i.e., $\vec{x}(t+\Delta t)=\vec{x}(\tau)$. Hence, considering each component of $\vec{x}$ as a particle in $1d$, the resetting event occurs simultaneously for all
the $N$ particles and they choose the same time $\tau$ in the past. A discrete-time version of this model of resetting with memory for $N=1$ particle was introduced and studied in Ref.~\cite{BS2014}. The generalisation to continuous-time
was studied in Ref.~\cite{BEM2017}. Here we further generalise this continuous-time model to $N>1$. 
Our goal is to study how dynamical correlations
between the coordinates of $\vec{x}$ emerge when resetting 
involves memory, i.e., a function $K_t(\tau)$ which is not simply $\delta(\tau)$ that characterises the kernel 
associated with memory-less resetting to the origin.

Let $P_r(\vec x,t)$ denote the joint PDF of the components of
$\vec x$ at time $t$ under resetting with memory. This quantity 
evolves via the Fokker-Planck equation
\begin{equation}
\partial_t P_r(\vec x,t)= D\, \nabla^2 P_r(\vec x,t)  - r\, P_r(\vec x,t) +r\, \int_0^t d\tau \int_{-\infty}^{\infty} d {\vec x\,'}\, K_t(\tau)\, P_r^{(2)}(\vec x\,',t; \vec x, \tau)\, ,
\label{fpr.1}
\end{equation}
starting from the initial condition $P_r(\vec x, t=0)= \delta(\vec x)$.
The first term on the right-hand-side (rhs) just represents diffusion and $D$ is the diffusion constant. The second term
represents the loss from position $\vec x$ due to simultaneous resetting with resetting rate $r$. The third term represents the gain term
to position $\vec x$ via resetting from a position $\vec x\,'$ reached at time $t$, where
$P_r^{(2)}(\vec x\,',t; \vec x, \tau)$ represents the joint probability density that the system
was at $\vec x$ at time $\tau\le t$ and at $\vec x\,'$ at time $t$.
We need to integrate over all the times $\tau$ that can be chosen by the particle, and over all the positions $\vec x\,'$ from which the particle may arrive at $\vec x$ by resetting. The reason for the solvability for the one-time marginal position distribution $P_r(\vec x, t)$ in this model can then be traced back to the fact that
\begin{equation}
\int P_r^{(2)}(\vec x\,', t; \vec x, \tau)\, d\vec x\,'= P_r(\vec x, \tau)\, ,
\label{marg.1}
\end{equation}
leading to a closed equation for the one-time PDF
\begin{equation}
\partial_t P_r(\vec x,t)= D\, \nabla^2 P_r(\vec x,t)  - r\, P_r(\vec x,t) +
r\, \int_0^t d\tau\, K_t(\tau) P_r(\vec x, \tau)\, .
\label{fpr.2}
\end{equation}
We next assume that $K_t(\tau)$ takes the form,
\begin{equation}
K_t(\tau)=\frac{\phi(\tau)}{\int_0^{t}dt'\, \phi(t')},
\label{Kgen}
\end{equation}
with $\phi(\tau)$ a non-negative function, and where the denominator in Eq. (\ref{Kgen}) ensures normalization, i.e., $\int_0^td\tau\, K_t(\tau)=1$. In the following, we will consider an exponential kernel,
\begin{equation}
\phi(\tau)= e^{-\lambda \tau},
\label{phiexp}
\end{equation}
with $\lambda$ a non-negative constant. Hence the walker tends to privilege the positions occupied at early times for resetting. This model interpolates between two well-studied cases.

When $\lambda\to\infty$, the particle resets to the initial position only. In this case $K_t(\tau)=\delta(\tau)$ and Eq. (\ref{fpr.2}) becomes
\begin{equation}
\partial_t P_r(\vec x,t)= D\, \nabla^2 P_r(\vec x,t)  - r\, P_r(\vec x,t) +
r\, \delta({\vec{x}}) .
\label{fpreset0}
\end{equation}
At late times, $P_r(\vec{x},t)$ approaches a NESS, $P^*(\vec{x})$, whose expression in $N$ 
dimension is given in Refs. \cite{EM2011,EM2014}.

When $\lambda\to0$, the random variable $\tau$ is distributed uniformly in the interval $[0,t]$, or $K_t(\tau)=1/t$. This case corresponds to the preferential relocation model \cite{BS2014,BEM2017} (or monkey walk \cite{MU2019}), where the probability density for choosing a position $\vec{x}$ for revisit is proportional to the local time spent by the walker at that site since the beginning of the process. Hence, more frequently visited regions of space are more likely to be revisited again. The Fokker-Planck equation (\ref{fpr.2}) becomes
\begin{equation}
\partial_t P_r(\vec x,t)= D\, \nabla^2 P_r(\vec x,t)  - r\, P_r(\vec x,t) +
\frac{r}{t}\, \int_0^t d\tau\ P_r(\vec x, \tau)\, .
\label{fprmonkey}
\end{equation}
This process has long range memory and is therefore highly non-Markov. At late times, as recalled further in Section \ref{sec:lambda0}, the  joint PDF above does not approach a NESS and remains time-dependent. The variance $\langle\vec x^2\rangle(t)$ keeps growing unbounded although very slowly, as it is proportional to $\ln(rt)$ at large time. 

Our goal is to compute the correlation coefficient $a_{\lambda}(t)$ given by Eq. (\ref{ac2c4}) which now depends on
$\lambda$, $t$ and $r$ (which is kept implicit in the notation).
In the next sub-sections, we show that $a_{\lambda}(t)$  can be obtained 
exactly at all times in these two limiting cases, namely $\lambda\to \infty$ and $\lambda\to 0$.

\subsection{Resetting to the origin, i.e., the limit $\lambda\to\infty$}
\label{sec:lambdainfty}

In the limit $\lambda\to \infty $, one obtains $K_t(\tau)=\delta(\tau) $ and this case simply corresponds to
the memory-less simultaneous resetting of all the components to the origin. In this limiting case,
the matrix elements $A_{ij}(t)$ in Eq. (\ref{corr_2.1}) were computed exactly for all $t$ in Ref.~\cite{DBMS2025}. We
reproduce below this calculation for the sake of completeness and compute the correlation coefficient $a_{\lambda\to\infty}(t)\equiv a_{\infty}(t)$
explicitly.

\begin{figure}[t]
\includegraphics[width=0.44\textwidth]{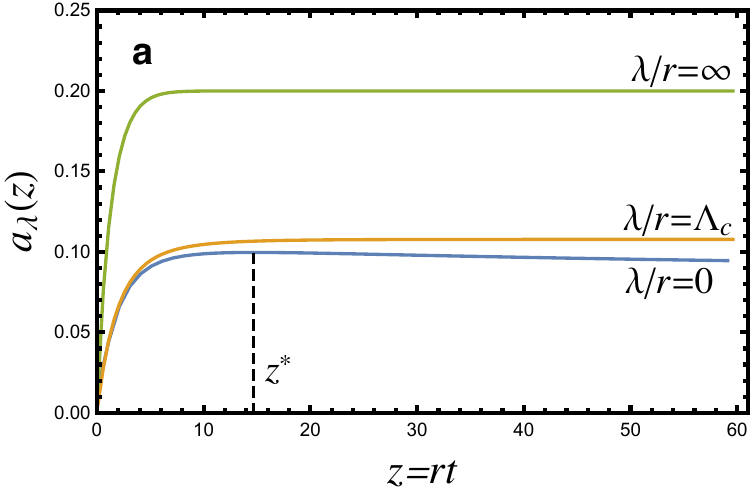}\hspace{0.3cm} 
\includegraphics[width=0.45\textwidth]{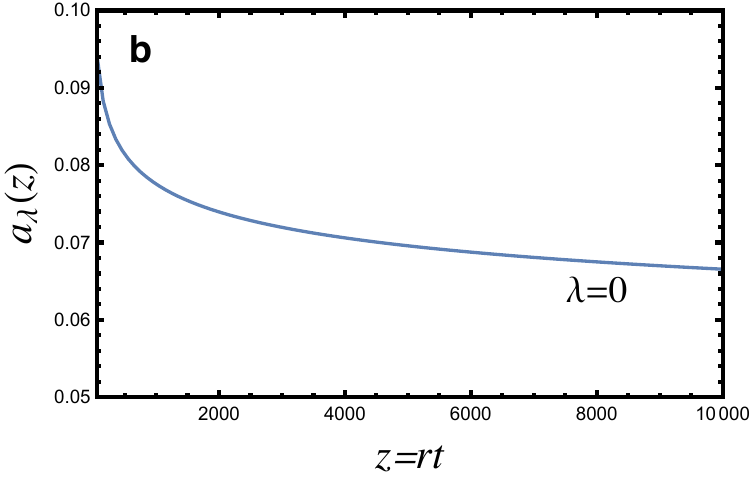} 
\caption{(a) Correlation coefficient $a_{\lambda}(z)$ defined in Eq. (\ref{ac2c4}) as a function of the rescaled time $z=rt$, for various particular values of $\lambda$. The case $\lambda=\infty$ (resetting to the origin) is given by Eq. (\ref{alambdainfty}) and tends exponentially fast to the asymptotic value $1/5$. In the case $\lambda=0$ (preferential relocation model), $a_0(z)$ is obtained from Eqs. (\ref{var.1b})-(\ref{num.1b}) and exhibits a non-monotonous behaviour with a maximum at $z^*=14.6732\ldots$. The correlation $a_0(z)$ increases linearly at small $z$ and, as shown by Panel (b) on a different scale of values of $z$, decays very slowly to $0$ at large $z$, as $1/\ln z$ to leading order. For all $\lambda>0$, $a_{\lambda}(z)$ tends to finite value as $z\to\infty$, which depends on $\lambda/r$ only and is given by Eqs. (\ref{azinfty})-(\ref{fy}). For $\lambda/r <\Lambda_c=0.0743\ldots $, $a_{\lambda}(z)$ has a maximum at a finite $z^*(\lambda/r)$, whereas for $\lambda/r\ge\Lambda_c$, $a_{\lambda}(z)$ increases monotonously with $z$ like in the case $\lambda=\infty$.}
\label{az.fig}
\end{figure}

Instead of solving Eq. (\ref{fpreset0}), it is more convenient to write directly the solution $P_r(\vec x,t)$ in the form of a renewal process \cite{EMS2020},
\begin{equation}
P_r(\vec x,t)=e^{-rt} G_0(\vec x,t)+r\int_0^t dt'\,e^{-rt'}G_0(\vec x,t'),
\label{renewal1}
\end{equation}
with $G_0(\vec x,t)$ the propagator of free diffusion in $N$ dimensions,
\begin{equation}
G_0(\vec x,t)=\frac{1}{(4\pi D t)^{\frac{N}{2}}}\,\exp\left(-\frac{\vec x^2}{4Dt}\right)=\prod_{i=1}^{N}\frac{1}{\sqrt{4\pi Dt}}\, e^{-\frac{x_i^2}{4Dt}} \,.
\end{equation}
The first term in the rhs of Eq. (\ref{renewal1}) counts the trajectories that reached $\vec x$ by diffusing freely because they never reset up to time $t$. The second term accounts for the trajectories that last restarted 
from the origin at a time $t-t'>0$ and then diffused freely from there until time $t$ without further resetting. Taking the Fourier transform of Eq. (\ref{renewal1}), one obtains
\begin{equation}
\tilde P_r(\vec k,t)= e^{-(r+Dk^2)t}+  r \frac{1- e^{-(r+Dk^2) t}}{r+Dk^2}\,,
\end{equation}
The expansion of the above expression up to order $k^4$ gives, in the notation of Eq. (\ref{Pkexpand}),
\begin{eqnarray}
c_2(z)&=&\frac{D}{r}\left(1-e^{-z}\right)\\
c_4(z)&=&\frac{D^2}{r^2}\left[1- (1+z)\,e^{-z}\right]\,, 
\end{eqnarray}
where we have defined the rescaled time variable,
\begin{equation}
z\equiv rt\, .
\end{equation}
Therefore the correlation coefficient $a_{\infty}(t)$ defined in Eq. (\ref{corrcoeff2}) is obtained from Eq. (\ref{ac2c4}) and only depends on $z$,
\begin{equation}
a_{\infty}(z)=\frac{1-2ze^{-z}-e^{-2z}}{5-(4+6z)e^{-z}-e^{-2z}}\,.
\label{alambdainfty}
\end{equation}
We deduce $a_{\infty}(z\to\infty)=1/5>0$, hence the correlation between the components (or particles) $x_1(t)$ and $x_2(t)$ persists in the steady state. The asymptotic behaviours are given by,
\begin{equation}
a_{\infty}(z)=\frac{1}{5}-\frac{4}{25}z e^{-z}+{\cal O}(e^{-z})\quad {\rm as}\quad z\to\infty\,,
\end{equation}
and
\begin{equation}
a_{\infty}(z)=\frac{z}{6}-\frac{z^2}{12}+{\cal O}(z^3) \quad\quad\quad {\rm as}\quad z\to0\,.
\end{equation}
The variations of $a_{\infty}(z)$ with $z$ are displayed in Fig. \ref{az.fig}a (top curve). The correlation coefficient is monotonously increasing, illustrating the building of dynamical correlations over time through resetting to the origin.

\subsection{Resetting with preferential relocation, or limit $\lambda\to0$}\label{sec:lambda0}

In this case, we start from the Fourier transform of Eq. (\ref{fprmonkey}),
\begin{equation}
\partial_t \tilde{P}_r(\vec k, t)= - (r+ D k^2) \tilde{P}_r(\vec k, t)
+ \frac{r}{t}\int_0^t \tilde{P}_r(\vec k, \tau)\, d\tau\, 
\label{FT_fp.1}
\end{equation}
where we recall that $k^2= k_1^2+k_2^2+\ldots +k_N^2$. 
In fact, Eq. (\ref{FT_fp.1}) is a straightforward generalization of the $N=1$ case 
(only $k_1^2$ gets replaced by $k^2$). Using the exact solution for $N=1$ given in Ref.~\cite{BEM2017}
and replacing $k_1^2$ by $k^2$ provides the exact solution for the $N$-particle case,
\begin{equation}
\tilde{P}_r(\vec k,t)= M\left(\frac{Dk^2}{r+Dk^2}, 1, - (r+Dk^2) t\right)\, ,
\label{sol.1}
\end{equation}
where $M(b,c,u)$ is the Kummer's confluent hypergeometric function defined
 by the power series
\begin{equation}
M(b,d,u)= \sum_{n=0}^{\infty} \frac{(b)_n\, u^n}{(d)_n\, n!}\, .
\label{Kummer.1}
\end{equation}
Here $(b)_n= b(b+1)(b+2)\ldots (b+n-1)$ is the Pochhammer symbol for $n\ge 1$, with $(b)_0=1$.

The exact solution (\ref{sol.1}) clearly demonstrates that at any finite time $t$,
the particle positions
are correlated, too, because $\tilde{P}(\vec k,t)$ does not factorise into a product
of indepedent marginal distributions. Consequently, in real space,
the joint distribution does not factorise,
\begin{equation}
P_r(\vec x, t) \ne p_r(x_1,t) p_r(x_2,t)\ldots p_r(x_N,t)\, ,
\label{corr.1}
\end{equation}
where $p_r(x_1,t) = \int P_r(x_1, x_2,\ldots, x_N, t) dx_2\,dx_3\ldots dx_N$ is the
marginal distribution of the first particle and so on. Clearly, this correlation is
induced dynamically via simultaneous resetting in the presence of memory.

Before calculating the correlation coefficient $a_{\lambda=0}(t)\equiv a_0(t)$ for all times, let us first investigate the limit of very large times $t\to \infty$.
In this case, using the asymptotic behaviour \cite{AS1965}, 
\begin{equation}
M(b,d, u) \approx \frac{\Gamma(d)}{\Gamma(d-b)}\, (-u)^{-b} \quad {\rm as}\quad u\to -\infty\, ,
\label{asymp.1}
\end{equation}
in Eq. (\ref{sol.1}), we get, to leading order for large $t$,
\begin{equation}
\tilde{P}_r(\vec k,t) \approx \frac{1}{\Gamma\left(\frac{r}{r+Dk^2}\right)}\, 
e^{-\frac{Dk^2}{r+Dk^2}\,\ln [(r+Dk^2)t]}\, .
\label{asymp.2}
\end{equation}
%\textcolor{blue}{
Expanding Eq. (\ref{asymp.2}) in powers of $k^2$ reveals that, at large $t$, the inverse transform will be dominated by the small $k$ regime.
%}
Taking the small $k$ and large $t$ limits, keeping $k^2 \ln (rt)$ fixed, we get
the asymptotic (long-time) scaling behaviour
\begin{equation}
\tilde{P}_r(\vec k,t) \approx e^{- \frac{D}{r}\ln (rt)\, k^2}= \prod_{i=1}^N 
e^{-\frac{D}{r}\, \ln(rt)\, k_i^2} \, .
\label{scaling_lim.1}
\end{equation}
Thus, in this scaling limit, the Fourier transform of the joint PDF
does indeed factorise. By inverting the Fourier transform, the joint PDF in the real space just becomes a product of Gaussians,
\begin{equation}
P(\vec x, t) \approx \prod_{i=1}^N \frac{1}{\sqrt{2\pi \sigma^2(t)}}\, 
e^{-x_i^2/{2\sigma^2(t)}} \quad {\rm where}\quad \sigma^2(t) = \frac{2D}{r}\, \ln (rt)\, .
\label{asymp_real.1}
\end{equation}
Therefore, in the long time scaling limit, the variables
$x_i(t)$ become statistically independent of each other, implying $a_0(t)\to0$ as $t\to\infty$. In addition, the distribution of each variable converges to a Gaussian with zero mean and variance $\sigma^2(t)$ growing logarithmically with time. Note that the variance of $\vec x(t)$ is obtained by summing over the $N$ components, or $\langle \vec x^2(t)\rangle \simeq \frac{2DN}{r}\ln(rt)$.
Clearly, there is no correlation between the components of $\vec x(t)$ at short times neither, as they are essentially independent Brownian motions. So, the correlation $a_0(t)$ must
behave non-monotonically with $t$.

To extract the coefficients $c_2(t)$ and $c_4(t)$ of the small $k$ expansion (\ref{Pkexpand}) at any time, it is useful to expand Eq. \eqref{sol.1} in power series
using Eq. \eqref{Kummer.1}. This gives, after small rearrangements,
\begin{equation}
\tilde{P}_r(\vec k,t)= 1+Dk^2\, \sum_{n=1}^{\infty} 
(2\, D\,k^2+r) (3 D\,k^2+2 r)\ldots (n\, D\,k^2+ r(n-1))\, \frac{(-t)^n}{(n!)^2}\, .
\label{series_exp.1}
\end{equation}
Expanding up to order ${\cal O}(k^4)$ is simple to do. Indeed, it is easy to see that for $n\ge 2$,
\begin{equation}
(2\, Dk^2+r) (3 Dk^2+2 r)\ldots (n\, D\, k^2+ r(n-1))= 
(n-1)!\left[ r^{n-1} + r^{n-2} (n-1+H_{n-1}) Dk^2 + O(k^4)\right]\, ,
\label{series_exp.2}
\end{equation}
where $H_n= 1+1/2+1/3+\ldots+ 1/n$ is the harmonic number. For $n=1$ in Eq. (\ref{series_exp.1}), the coefficient of $-t$ is simply
$Dk^2$. Putting this result back into \eqref{series_exp.1} gives
\begin{equation}
\tilde{P}_r(\vec k,t) = 1+\frac{Dk^2}{r}\, \sum_{n=1}^{\infty} \left[1+ (n-1+H_{n-1})\, \frac{Dk^2}{r}+ {\cal O}(k^4)\right] \frac{(-rt)^n}{n\, n!}\, ,
\label{series_exp.3}
\end{equation}
where we interpret $H_0=0$. Using the rescaled time $z=rt$, we deduce
\begin{equation}
c_2(z)=-\frac{D}{r} \sum_{n=1}^{\infty}\frac{(-z)^n}{n\, n!}\, ,
\label{var.1}
\end{equation}
and 
\begin{equation}
c_4(z)=\frac{D^2}{r^2}\sum_{n=2}^{\infty} (n-1+H_{n-1}) \frac{(-z)^n}{n\, n!}\, .
\label{num.1}
\end{equation}
The two above sums can be carried out by using Mathematica and can be rewritten as,
\begin{equation}
c_2(z)=\frac{D}{r}\left[E_1(z)+ \gamma_E+ \ln z\right]\, ,
\label{var.1b}
\end{equation}
and
\begin{equation}
c_4(z)=\frac{D^2}{r^2}\, \left[ \gamma_E -1 + E_1(z) + \ln z+ e^{-z}+ \frac{1}{2}\,
_1F_1^{(2,0,0)}(0,1,-z)\right]\, ,
\label{num.1b}
\end{equation}
%Now, evaluating \eqref{small_k.1}, we immediately get for $i\ne j$
%\begin{equation}
%\langle x_i^2(t) x_j^2(t)\rangle= \frac{8 D^2}{r^2}\sum_{n=2}^{\infty} (n-1+H_{n-1}) \frac{(-rt)^n}{n\, n!}\, =\frac{8 D^2}{r^2}\, \left[-1+ \gamma_E + E_1(rt) + \ln (rt)+ e^{-rt}+ \frac{1}{2}\, _1F_1^{(2,0,0)}(0,1,-rt)\right]\, ,
%\label{num.1}
%\end{equation}
where $\gamma_E=0.5772\ldots$ is the Euler constant, $E_1(z)= 
\int_z^{\infty} \frac{e^{-x}}{x}\, dx$
and $_1F_1^{(2,0,0)}(0,1,-z)= \partial_b^2 M(b,1,-z)\Big|_{b=0}$ with $M(b,d,u)$ defined 
in Eq. (\ref{Kummer.1}). 
%The variance $\langle x_i^2(t)\rangle$ can also be easily extracted from \eqref{series_exp.3} using the formula
%\begin{equation}
%\langle x_i^2(t)\rangle=- \frac{\partial^2 \tilde{P}_r(\vec k, t)}{\partial k_i^2}\Big|_{\vec k=\vec 0} = -\frac{2D}{r} \sum_{n=1}^{\infty} \frac{(-rt)^n}{n\, n!}=\frac{2D}{r}\left[E_1(rt)+ \gamma_E+ \ln (rt)\right]\, .
%\label{var.1}
%\end{equation}
Substituting \eqref{var.1b} and \eqref{num.1b} in the expression (\ref{ac2c4}) gives the sought coefficient $a_0(z)$, which also only depends on $z$.
%\begin{equation}
%a(t) =  \frac{2\, \left[-1+ \gamma_E + E_1(rt) + \ln (rt)+ e^{-rt}+ \frac{1}{2}\, _1F_1^{(2,0,0)}(0,1,-rt)\right]}{ \left[E_1(rt)+ \gamma_E+ \ln (rt)\right]^2}-1\, .
%\label{At_final.1}
%\end{equation}
The function $a_0(z)$ is plotted in Fig. \ref{az.fig}a (bottom curve). Numerically, we find that the correlation is maximal at $z=z^*$ with
\begin{equation}
 z^*=14.6732\ldots\, .
 \label{zstar}
\end{equation}
And at the maximum, 
\begin{equation}
a_0(z^*)= 0.0995506\ldots\, ,
\end{equation}
indicating a weaker correlation than in the previous case of resetting to the origin. The asymptotic behaviours are,
\begin{eqnarray}
\label{At_asymp.1}
a_0(z)= \begin{cases}
& \frac{1}{18}\, z  +O\left(z^2\right) \quad\quad\,\,\, {\rm as}\quad z\to 0 \\
\\
& \frac{1}{\ln z}+ O\left(\frac{1}{\ln^2 z}\right) \quad\quad {\rm as}\quad z\to \infty \, .
\end{cases}
\end{eqnarray}
The correlation thus tends extremely slowly to 0 at large $z$, as illustrated by Fig. \ref{az.fig}b on a different scale, meaning that the components remain correlated during a very long time in practice.
While the small $t$ behaviour is easy to extract, computing the large $t$ asymptotics is a bit tricky since Mathematica is unable to provide it. However the precise asymptotics for large $t$, leading to the second line
in Eq. (\ref{At_asymp.1}),  is derived in Appendix \ref{sec:appA}.

\section{Correlation in the general case $0<\lambda<\infty$}\label{sec:anylambda}

We now consider the exponential memory kernel (\ref{phiexp}) for any $\lambda>0$. The Fourier transform of Eq. (\ref{fpr.2}) reads
\begin{equation}
\partial_t \tilde{P}_r(\vec k, t)= - (r+ D k^2) \tilde{P}_r(\vec k, t)
+ \frac{\lambda r}{1-e^{-\lambda t}}\int_0^t e^{-\lambda \tau}\tilde{P}_r(\vec k, \tau)\, d\tau\,. 
\label{FT_fp.exp}
\end{equation}
As for the preferential relocation model, Eq. (\ref{FT_fp.exp}) was exactly solved for the $N=1$ dimensional case in Ref. \cite{BEM2017}. We can then use that solution, replacing $k_1^2$ by $k^2$ to obtain the solution in $N$ dimensions,
\begin{eqnarray}
\tilde{P}_r(\vec k, t) &=& \frac{\Gamma(1-C)}{\Gamma(1-A)\Gamma(1-B)} F(A,B,C;\, e^{-\lambda t})\nonumber\\
&&+\frac{1}{1-C}\frac{\Gamma(C)}{\Gamma(A)\Gamma(B)}\,e^{-\lambda(1-C)t}
F(1-B,1-A,2-C;\, e^{-\lambda t})\,,
\label{sollambda}
\end{eqnarray}
where $F(u,v,w;\, \chi)$ is the standard hypergeometric function,
\begin{equation}
F(u,v,w;\, \chi)=\sum_{n=0}^{\infty}\frac{(u)_n(v)_n}{(w)_n}\frac{\chi^n}{n!}\, ,
\label{2f1}
\end{equation}
and
\begin{eqnarray}
C&=&1-\frac{r+Dk^2}{\lambda}\label{c}\\
A&=&\frac{1}{2}\left( C+\sqrt{C^2+\frac{4Dk^2}{\lambda}} \right)\label{b}\\
B&=&\frac{1}{2}\left( C-\sqrt{C^2+\frac{4Dk^2}{\lambda}} \right)\, . \label{a}
\end{eqnarray}

\subsection{Infinite time limit}\label{sec:lamb_infinite_t}

\begin{figure}[t]
\includegraphics[width=0.5\textwidth]{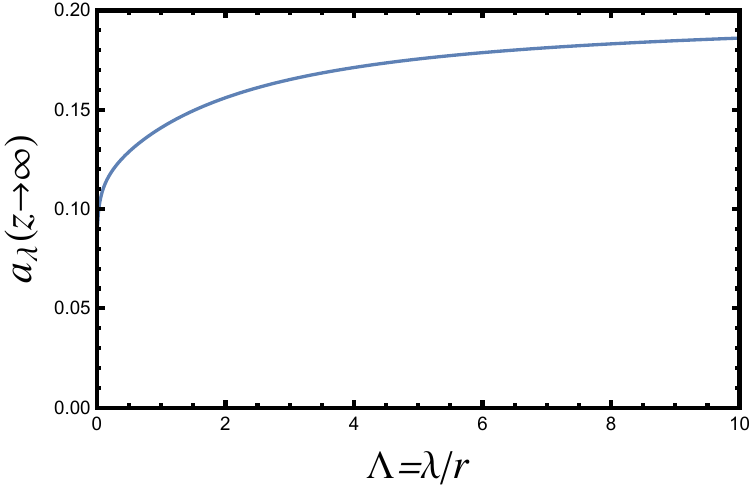} 
\caption{Asymptotic value of the correlation at large times as a function of $\Lambda=\lambda/r$.}
\label{zinfty.fig}
\end{figure}

We first analyse the correlation $a_{\lambda}(z)$ in the limit $z=rt\to\infty$, which is more tractable than the finite time case. At late times, $\tilde{P}_r(\vec{k},t)$ in Eq. (\ref{sollambda}) tends to a NESS given by,
\begin{equation}
\tilde P_r^{\rm (st)}(\vec k)= \frac{\Gamma(1-C)}{\Gamma(1-A)\Gamma(1-B)}      \, ,
\label{nesslambda}
\end{equation}
for all $\lambda>0$ and $r>0$. By expanding the expression (\ref{nesslambda}) up to order $k^4$ and using Eq. (\ref{ac2c4}), one finds that the correlation at infinite time is non-zero, like in the case of resetting to the origin $\lambda=\infty$, and that it is a function of $r/\lambda$ only, 
\begin{equation}
a_{\lambda}(z\to\infty)= f\left(\frac{r}{\lambda}\right)\, ,
\label{azinfty}
\end{equation}
with
\begin{equation}
\frac{1}{f(y)}=3-\frac{12(y-1)[\gamma_E+\psi(y)]^2}{(y-1)[\pi^2+6(2y-1)\psi'(y)]-12y[\gamma_E+\psi(y)]}\, ,
\label{fy}
\end{equation}
where $\psi(y)=\Gamma'(y)/\Gamma(y)$ is the polygamma function of 0-th order. The function $a_{\lambda}(z\to\infty)$ has the following asymptotic behaviours with respect to the variable $\lambda/r$:
\begin{eqnarray}
\label{alambda_asymp.1}
a_\lambda(z\to\infty)= \begin{cases}
& -\frac{1}{\ln (\lambda/r)}\,+\, ... \quad\quad\quad {\rm as}\quad \lambda/r\to 0 \, \\
\\
& \frac{1}{5}-\frac{4}{25}\left(\frac{\lambda}{r}\right)^{-1}\, +\,... \quad {\rm as}\quad \lambda/r\to \infty\,.
\end{cases}
\end{eqnarray}
Hence the value of the correlation $a_{\lambda}(z\to\infty)$ at large times is a monotonically 
increasing function of the rescaled memory parameter $\Lambda\equiv\lambda/r$.
The asymptotic correlation grows from $0$ to $1/5$, as shown by Fig. \ref{zinfty.fig}. Due to the logarithmic singularity in the first line of Eq. (\ref{alambda_asymp.1}), as soon as $\Lambda\sim10^{-2}$, the asymptotic correlation is quite large already ($\sim 0.1$).

\subsection{Correlation at finite time}\label{sec:lamb_finite_t}

This case is the most general here, as the correlation $a_{\lambda}(t)$ depends on the three variables 
$\lambda$, $r$ and $t$ [the diffusion constant $D$ cancels
out between the numerator and denominator of Eq. (\ref{ac2c4})]. Using the $\Pi-$theorem of dimensional analysis \cite{B1996} and noticing that $a_{\lambda}$ is dimensionless, the correlation must depend on two dimensionless variables, for instance,
\begin{equation}
a_{\lambda}(t)=\alpha(rt,\lambda/r)=\alpha(z,\Lambda)\, ,
\label{pitheorem}
\end{equation}
with $\alpha$ a function. The above relation actually allows us to check {\em a posteriori} that in the 
limits $\lambda\to 0$ or $\lambda\to \infty$, the correlation only depends on $z=rt$. 
Conversely, for a finite $\lambda$ but $t\to\infty$, it only depends on $\Lambda=\lambda/r$
given explicitly in Eqs. (\ref{azinfty}) and (\ref{fy}).
Extracting the coefficients $c_2(t)$ and $c_4(t)$ from Eq. (\ref{sollambda}) at finite $t$ is cumbersome and we have opted to perform the Taylor expansion in powers of $k^2$ numerically instead. Setting $K\equiv k^2$, the coefficients are obtained from the relations $c_2(t)=-\partial \tilde P_r(K,t)/\partial K|_{K=0}$ and
$c_4(t)=\frac{1}{2!}\,\partial^2 \tilde P_r(K,t)/\partial K^2|_{K=0}\,$, where the derivatives of the solution (\ref{sollambda})  are evaluated numerically with Mathematica.

\begin{figure}[t]
\includegraphics[width=0.5\textwidth]{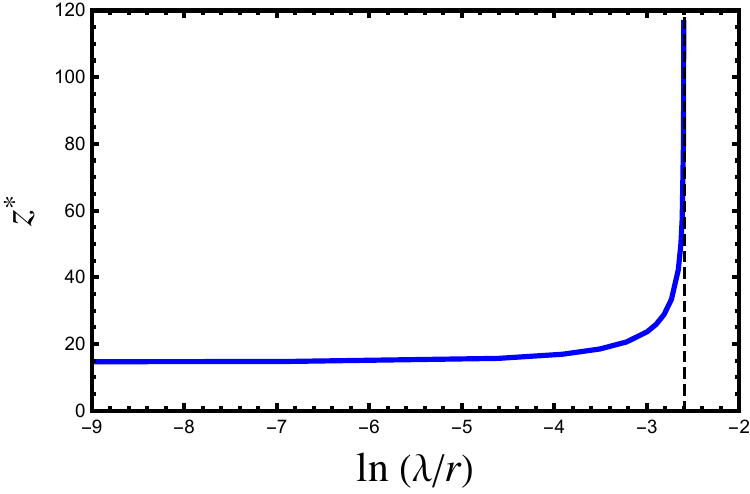} 
\caption{Rescaled time $z^*$ where the correlation $a_{\lambda}(z)$ is maximal, 
depicted in Fig. \ref{az.fig}, as a function of $\lambda/r$ (in log scale). 
For $\lambda/r<\Lambda_c=0.0743\ldots$, $z^*$ is finite and $a_{\lambda}(z)$ non-monotonous. 
For $\lambda/r>\Lambda_c$, or a memory kernel sufficiently peaked at short times, 
$a_{\lambda}(z)$ increases monotonously with $z$ and the maximal correlation is reached at infinite times ($z^*=\infty$).
The vertical dashed line shows the location of the critical value 
$\ln(\Lambda_c)=\ln(0.0743\cdots)=-2.59964\ldots$.}
\label{zstar.fig}
\end{figure}

Figure \ref{az.fig}a shows the obtained correlation $a_{\lambda}(z)$ as a function of $z$, fixing $\Lambda=0.0743\ldots$ (middle curve). We find that this value is peculiar, in the sense that it separates two different regimes.
For $0\le\Lambda<0.0743\ldots$, the function $\alpha(z,\Lambda)$ is non-monotonous with $z$ and exhibits a maximum at a certain time $z^*(\Lambda)$, like in the case $\lambda=0$. For $\Lambda>0.0743\ldots$, however, $\alpha(z,\Lambda)$ is a monotonous function of $z$ and reaches its maximum at $z=\infty$, like for $\lambda=\infty$. In Fig. \ref{zstar.fig}, we display $z^*$ as a function of $\Lambda$. Remarkably, $z^*$ increases with $\Lambda$ and diverges when $\Lambda\to\Lambda_c=0.0743\ldots$ from below. 

\section{Renewal-like equation for resetting with memory}\label{sec:renewallike}

In this section we show that the emergence of correlations in processes with simultaneous 
resetting to the initial position or to sites visited in the past can be described in a unified way 
by  noticing their connection with conditionally independent and identically distributed 
(c.i.i.d.) random variables \cite{BLMS2023, BLMS2024, BKMS2024,SM2024,MMS2025,DBMS2025,KMS2024,MKMS2025}.
Consider first the case $\lambda\to \infty$ where a single Brownian walker in $N$ dimensions
resets to the origin with rate $r$, or equivalently $N$ independent one-dimensional Brownian
motions reset simultaneously to the origin with rate $r$.
The renewal equation (\ref{renewal1}) describes the joint PDF of the $N$ components of the
position $\vec x$ of the single walker at time $t$. 
This equation (\ref{renewal1}) can be expressed in a more convenient form,
\begin{equation}
P_{r}(\vec x,t)= \int_0^{\infty} dt'\ \varphi_t(t')\, \prod_{i=1}^N\, p_0(x_i,t')\, ,
\label{ciid1}
\end{equation}
where $\varphi_t(t')=e^{-r t'}\delta(t'-t)+re^{-rt'}$ for $t'\le t$ and 
$\varphi_t(t')=0$ for $t'>t$, while $p_0(x_i,t')= G_0(x_i,t')= e^{-x_i^2/{4Dt'}}/\sqrt{4\pi\,D\,  t'}$ 
is the propagator of a free Brownian motion in $1d$. 
The function $\varphi_t(t')$ represents the PDF of the time $t'$ elapsed between the latest restart and 
the current time $t$. Conditioned on $t'$, the variables $x_i$'s are actually 
uncorrelated as indicated by the product of the $p_0$'s, but the presence of the 
integral over $t'$ correlates them. Consequently, the joint PDF $P_r(\vec x,t)$ cannot be written as a 
product of functions of one variable $x_i$. The structure given by Eq. (\ref{ciid1}) defines 
conditionally i.i.d. variables more generally, where $t'$ can be seen as a random parameter with a certain 
distribution that may or may not depend on $t$ \cite{BLMS2024}.

\begin{figure}[t]
\includegraphics[width=0.5\textwidth]{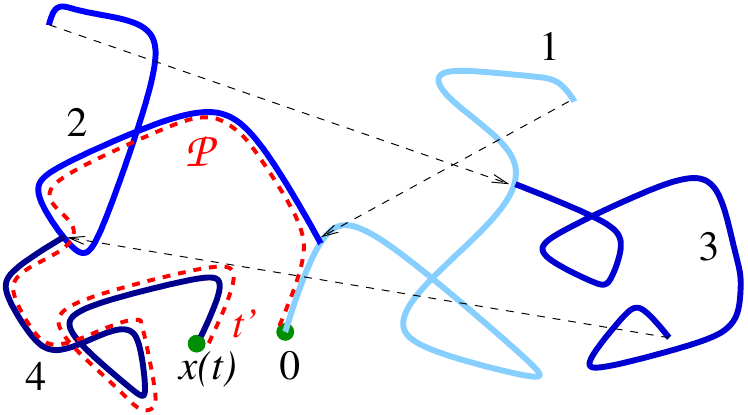} 
\caption{Illustration of a two-dimensional diffusion process after 3 resetting events to previous positions, or 4 consecutive paths. The long-range jumps to previous points of the trajectory are shown by dashed arrows. The position $\vec x(t)$ of the particle at time $t$ can also be reached by following the continuous Brownian path ${\cal P}$ (in red), which is of much shorter duration $t'$. The components of the position $\vec x(t)$ have a c.i.i.d. structure.}
\label{ciid.fig}
\end{figure}

We now turn to the case of resetting with memory and a general kernel. 
It turns out that the joint distribution $P_r(\vec x,t)$ in this case
can also be written under an exact form with a structure very similar to Eq.~(\ref{renewal1}), namely,
\begin{equation}
 P_r(\vec x,t)=e^{-rt}G_0(\vec x,t)+\int_0^{\infty}dt'\, \Psi_t(t')\, G_0(\vec x,t')\, ,
\label{renewal2}
\end{equation}
or equivalently
\begin{equation}
P_r(\vec x,t)= \int_0^{\infty}dt'\, \Phi_t(t')\, \prod_{i=1}^N\, p_0(x_i,t')\, ,
\label{ciid2}
\end{equation}
with $\Phi_t(t')=e^{-r t'}\delta(t'-t)+\Psi_t(t')$ and $p_0(x_i,t')=G_0(x_i,t')=e^{-x_i^2/{4Dt'}}/\sqrt{4\pi\, D\, t'}$
denoting again the one dimensional free Brownian propagator. 
Eq. (\ref{ciid2}) has a  c.i.i.d. structure like Eq. (\ref{ciid1}). 
However, the time $t'$ has a quite different meaning and its PDF $\Phi_t(t')$ is no longer given 
by $\varphi_t(t')$. 
%\textcolor{blue}{
The function $\Psi_t(t')$ is the PDF of the variable $t'$ (to be defined below) due to the trajectories which had one or more resets at time $t$.
%}
The definition of $t'$ is illustrated in Fig. \ref{ciid.fig} with a two-dimensional process ($N=2$),
and can be explained as follows. 
The first term in the rhs of Eq. (\ref{renewal2}), represents, as in standard resetting, the probability that the 
particle has never reset up to time $t$ and has reached the position $\vec x$ by pure diffusion. This term is negligible at 
late times. In the case of one or multiple resetting events, one can notice that the position $\vec x(t)$ can be reached 
from the initial position $0$ by following a continuous path ${\cal P}$ free of resetting jumps between distant positions
(as shown by a red dashed line in Fig.~\ref{ciid.fig}). 
This is a consequence of the fact that resetting occurs to previously visited positions. In Fig. \ref{ciid.fig}, for instance, the 
particle performs a first Brownian path starting from the origin and resets to a previous position and
the path 2 starts from this location.
At the end of path 2, the chosen time in the past $\tau$ belongs to path 1 and the corresponding position is the beginning 
of path 3. Similarly, path 4 starts from a point belonging to path 2 and the particle arrives at $x(t)$ without further 
resetting. On this trajectory, a plain Brownian particle could have reached ${\vec x}(t)$ by taking a piece of path 1 (for a time 
$t'_1$), then a piece of path 2 (for a time $t'_2$), and finally all path 4 (of duration $t'_4$). This defines the 
continuous path ${\cal P}$ in red in Fig. \ref{ciid.fig}, while $t'$ is the duration of 
${\cal P}$, i.e., $t'=t'_1+t'_2+t'_4$ in our example. For any 
trajectory $\{\vec x(\tau)\}_{0\le \tau\le t}$ up to an arbitrary time $t$, there exists a unique 
path ${\cal P}$ and time $t'\le t$.

In standard resetting to the origin, the relevant part of the trajectory is
the most recent Brownian path starting from the origin following the last reset to the origin. But
for resetting with memory,  the relevant path is 
now ${\cal P}$, which also starts from the origin and is Brownian, since it is composed of a succession of independent 
Brownian motions. The quantity $t'$ now denotes the duration of ${\cal P}$, and not
the time since last resetting as in standard resetting to the origin. 
Having provided a clear physical meaning of the relevant time $t'$, we then note that
conditioned on $t'$, the variables $x_i$'s are distributed with $G_0(\vec x, t')$ and independent. 
These considerations lead to Eq. (\ref{ciid2}), where the PDF of $t'$ at time $t$ is $\Phi_t(t')$.

Thus, while the c.i.i.d structure of the joint distribution is manifest in Eq. (\ref{ciid2}),
the real non-trivial problem is to determine the distribution $\Phi_t(t')$ of $t'$
for a fixed $t$. Employing a mapping to weighted random recursive trees, 
Mailler and Uribe Bravo have been able to characterise $\Phi_t(t')$ rigorously at large $t$ 
for several long-range memory kernels and a wide class of resetting protocols \cite{MU2019}. 
In the case of a uniform memory kernel ($\lambda=0$) and when the time intervals between resetting events are 
exponentially distributed (the resetting rate $r$ is constant), their results stipulate that the rescaled variable
\begin{equation}
\xi=\frac{t'-\frac{1}{r}\ln(rt)}{\frac{1}{r}\sqrt{2\ln(rt)}}
\label{MU}
\end{equation}
becomes Gaussian distributed with zero mean and unit variance as $t\to\infty$. 
The general proof is carried out by applying the strong law of large numbers and central limits theorems on sums of 
independent and non-identically distributed random variables \cite{MU2019}. Hence $\Phi_t(t')$ is always time dependent, 
peaked around $t'^*=\frac{1}{r}\ln(rt)$ and with a much smaller standard deviation $\frac{1}{r}\sqrt{2\ln(rt)}$. These 
features contrast with the function $\varphi_t(t')$ in Eq. (\ref{ciid1}), which tends toward the stationary exponential 
distribution $\varphi_{\infty}(t')=r e^{-rt'}$ and whose mean and standard deviation are both equal to $1/r$. 
To leading order for large $t$, the limiting form in Eq. (\ref{MU}) suggests that $\Phi_t(t')$ can
be approximated by a delta function centered at the mean, i.e., $\Phi_t(t')\approx \delta(t'-t'^*)$
where $t'^*= \frac{1}{r}\ln(rt)$.
Substituting 
the limit form of $\Phi_t(t')$ into Eq. (\ref{ciid2}), one recovers the distribution $P_r(\vec x,t)$ at large $t$ as a 
product of Gaussians with a variance growing as $2\, D\, t'^*=\frac{2D}{r}\ln(rt) $, as shown by Eq. (\ref{asymp_real.1}) in 
Section \ref{sec:lambda0}.

The decay of correlations between the components of $\vec x(t)$ at late times for $\lambda=0$ is therefore a consequence of 
the growing characteristic time-scale $t'^*$ associated with the paths ${\cal P}$. If $\lambda>0$, in contrast, as 
discussed above, the joint distribution converges to a NESS. 
%and correlations persist as $t\to\infty$. 
The existence of a NESS implies that 
$\Phi_t(t')$ tends to a non-trivial stationary distribution $\Phi_{\infty}(t')$ in Eq. (\ref{ciid2}) when $t\to \infty$. As the particle mostly 
resets to positions visited early and belonging to the first diffusive paths $\{1,2,3,\ldots\}$, the paths ${\cal P}$ remain 
short even after a very large number of resetting events.
%\textcolor{blue}{
The standard deviation of $\Phi_{\infty}(t')$ is not negligible compared to its average in principle, therefore Eq. (\ref{ciid2})  cannot be simplified as a product of distributions. We deduce that the existence of a NESS and the persistence of correlations are closely related.
%}
This situation is similar to resetting to the origin and the 
results of Mailler and Uribe Bravo do not apply to these cases, as they have considered memory kernels $K_t(\tau)$ which did 
not decay rapidly with $\tau$.

Eq. (\ref{ciid2}) further allows us to establish a simple general relation between the quantities $P_r$ and $\Phi_t$ for Brownian motion, valid for any time $t$. 
In fact, for the preferential relocation model ($\lambda=0)$, the relation between these two quantities also provides us with an alternative derivation of the result (\ref{MU}) of~\cite{MU2019}, namely that the centred and scaled variable $\xi$ has a normal distribution of zero mean and unit variance.
Let us proceed as follows. We first take the Fourier transform of Eq. (\ref{ciid2}) that gives
\begin{equation}
\tilde{P}_r(\vec k, t)= \int_0^{\infty} dt'\, \Phi_t(t')\, \prod_{i=1}^N \tilde{p}_0(k_i,t')\, ,
\label{FT_full.1}
\end{equation}
where $\tilde{p}_0(k_i,t')= e^{-D\, k_i^2\, t'}$ is simply the Fourier transform of the propagator of a Brownian motion in 
one dimension at time $t'$. Using $k^2= \sum_{i=1}^N k_i^2$ then gives
\begin{equation}
\tilde{P}_r(\vec k, t)= \int_0^{\infty} dt'\, \Phi_t(t')\, e^{-D\, k^2\, t'}\, .
\label{FT_full.2}
\end{equation}
Thus if $\tilde{P}_r(\vec k, t)$ is known, we can invert this relation to determine $\Phi_t(t')$ in principle.
For the preferential relocation model ($\lambda=0$), we know $\tilde{P}_r(\vec k, t)$ exactly  
from Eq. (\ref{sol.1}). Substituting this result on the left-hand-side of (\ref{FT_full.2}), and replacing $ D\, k^2$ by a new variable $s$ for convenience, we get
\begin{equation}
\hat{\Phi}_t(s)=\int_0^{\infty} dt'\, \Phi_t(t')\, e^{-s\, t'}= 
M\left(\frac{s}{r+s}, 1, - (r+s) t\right)\, ,
\label{laplacePhi}
\end{equation}
where $\hat{\Phi}_t$ is the Laplace transform of $\Phi_t$. In the case $\lambda>0$, an explicit but long expression can similarly be obtained from Eq. (\ref{sollambda}).
Hence, for simplicity,
let us focus on the case $\lambda=0$ only. The Laplace transform
in Eq. (\ref{laplacePhi}) can be formally inverted to give
\begin{equation}
\Phi_t(t')= \int_{\Gamma} \frac{ds}{2\pi i}\, e^{s t'}\, M\left(\frac{s}{r+s}, 1, - (r+s) t\right)\, ,
\label{Brom.1}
\end{equation}
where $\Gamma$ denotes a vertical Bromwich contour in the complex $s$-plane whose real part lies to the right
of all singularities of the integrand. An explicit inversion using (\ref{Brom.1}), valid for all $t$, seems difficult.
However, one can make progress for large $t$, where we can replace the function $M$ by its asymptotic expansion,
corresponding to the limit $u\to -\infty$  in Eq. (\ref{asymp.1}). This gives, as $t\to \infty$ and fixed $s$,
\begin{equation}
M\left(\frac{s}{r+s}, 1, - (r+s) t\right)\approx \frac{1}{\Gamma\left(\frac{r}{r+s}\right)} e^{-\frac{s}{r+s}\, \ln [(r+s)t]}\, .
\label{M_asymp.1}
\end{equation}
Next, we want to investigate the behaviour of $\Phi_t(t')$ when both $t'$ and $t$ are large. For large $t'$, the
dominant contribution to the integral in Eq. (\ref{Brom.1}) comes from the vicinity of small $s$, as we can shift the
contour $\Gamma$ close to $s=0$. Hence we can now expand the integrand in Eq. (\ref{Brom.1}) in small $s$. Substituting
\begin{equation}
\frac{s}{r+s}\, \ln [(r+s)t]\approx \frac{s}{r}\, \ln(rt) - \frac{s^2}{r^2}[\ln (rt)-1]+ O(s^3)\, ,
\label{small_s.1}
\end{equation}
in Eq. (\ref{M_asymp.1}), we obtain from Eq. (\ref{Brom.1}), when $t$ and $t'$ are both large,
\begin{equation}
\Phi_t(t')\approx  \int_{\Gamma} \frac{ds}{2\pi i}\, e^{s [t'- \frac{1}{r}\, \ln (rt)] + \frac{s^2}{r^2}\, \ln (rt) +\ldots}\, ,
\label{Brom.2}
\end{equation}
where $\ldots$ denote higher order terms in $s$.  Rescaling
\begin{equation}
s= \frac{r}{\sqrt{2\, \ln(rt)}}\, \tilde{s}\, ,
\label{s_rescale.1}
\end{equation}
and identifying the variable $\xi$ in Eq. (\ref{MU}), we can re-write the integral in Eq. (\ref{Brom.2}) as
\begin{equation}
\Phi_t(t')\approx \frac{r}{\sqrt{2\ln (rt)}} \int_{\Gamma} \frac{d\tilde{s}}{2\pi i}\, e^{\tilde{s}\, \xi + \frac{1}{2}\, \tilde{s}^2}\, .
\label{Brom.3}
\end{equation}
Note that under the rescaling \eqref{s_rescale.1}, all higher order terms in small $s$, beyond the quadratic one inside the exponential in Eq. (\ref{Brom.2}),
get rescaled by positive powers of $1/\sqrt{\ln (rt)}$ and hence disappear for large $t$. Finally, the integral in
Eq. (\ref{Brom.3}) is simple to perform exactly as it is just a Gaussian integral. This brings us to the final
result that in the limit $t$ and $t'$ both large, but with the scaling combination $\xi$ in Eq. (\ref{MU}) held fixed,
the function $\Phi_t(t')$ approaches the scaling form
\begin{equation}
\Phi_t(t')\approx \frac{r}{\sqrt{2\ln (rt)}}\, f_0\left(\frac{t'-\frac{1}{r}\ln(rt)}{\frac{1}{r}\sqrt{2\ln(rt)}}\right)\, , \quad\quad
{\rm where} \quad f_0(\xi)= \frac{1}{\sqrt{2\pi}}\, e^{-\xi^2/2}\, .
\label{clt_l0}
\end{equation}
This completes the derivation of the result in the preferential relocation model ($\lambda=0$) that states that
the centred and scaled variable $\xi$ approaches a normal distribution with zero mean and unit variance at long times.

\section{Conclusions}\label{sec:concl}

We have studied the time evolution of the correlations between $N$ independent one-dimensional diffusive processes starting from the 
origin and subject to simultaneous resetting, in cases where resetting occurs to positions visited in the past. This 
problem is equivalent to the study of the mutual correlations between the components of the position of a single $N$ 
dimensional walker subject to resetting to positions visited in the past. One may consider, for instance, the 
$(x_1,x_2)$-coordinates of a random walker on the plane that revisits familiar places in its environment. We have 
chosen a memory kernel that interpolates, as a parameter $\lambda$ is varied, between resetting to the starting 
position only (a memory-less process) and the preferential relocation model, which is strongly non-Markov. Our study 
provides an exact calculation of the full time-dependent emergence of the correlations in a 
non-Markov resetting model with memory, thus extending previous studies~\cite{BLMS2023,BLMS2024,DBMS2025,BCKMPS2025} 
on the memory-less case.

For short-ranged memory kernels (large $\lambda$, similar to resetting to the origin) the correlation coefficient 
$a_{\lambda}(t)$ defined in Eq. (\ref{corrcoeff1}) monotonically increases from 0 with time. Conversely, when memory is 
sufficiently long-range, correlations exhibit a non-monotonic behaviour with time and reach a maximum at a certain re-scaled 
time $z^*$. These two regimes are separated by a critical value $\lambda_c$ of the parameter $\lambda$, where $z^*$ 
diverges. For any non-zero $\lambda$, $a_{\lambda}(t)$ tends to a finite asymptotic value at large $t$. This persistent 
correlation depends only on $\lambda$ and the resetting rate $r$ through Eqs. (\ref{azinfty})-(\ref{fy}). Therefore 
simultaneous resetting represents a robust mechanism by which strong correlations are built and maintained 
dynamically. The correlation is the largest in the case of resetting to the origin and $a_{\infty}(t\to\infty)=1/5$ in that 
case, in agreement with previous results \cite{BLMS2023,BLMS2024}.

The preferential relocation model (case $\lambda=0$) is peculiar because the joint PDF of $(x_1(t),x_2(t))$ does not 
tend to a NESS and keeps depending on time at large times. We find that $a_0(t)$ is non-monotonous as well, but tends to 
$0$ as $t\to\infty$. Hence the processes are uncorrelated at both short and large times. However, the decay of the 
correlations at late times is extremely slow, as $a_0(t)\simeq 1/\ln(rt)$.

We have finally shown that multi-dimensional processes subject to resetting (with or without memory) generally obey a 
conditionally i.i.d. structure at all times. This means that the joint distribution of the components of $\vec x(t)$ 
can be written under the form in Eq. (\ref{ciid2}). For instance, in the standard renewal approach of resetting to the 
origin, conditioning on the time $t'$ elapsed since the latest reset, the positions $x_1(t)$ and $x_2(t)$ are 
independent. When memory is present, the renewal approach no longer holds {\em per se} but the c.i.i.d. structure 
remains.  In this case, the time $t'$ is also the duration of a Brownian path that goes from the origin to the current 
position without any discontinuities, but it is formed by several pieces of previous Brownian paths, not just the 
latest one. Thus, the c.i.i.d structure of the joint distribution is not apparently manifest but remains {\em hidden} 
in this model, as in the case of another recently studied model of independent particles in a switching harmonic 
trap~\cite{BKMS2024}. The challenge is to find the hidden c.i.i.d. structure explicitly because such a structure then 
guarantees exact computation of several physical observables in a relatively straightforward manner~\cite{BKMS2024}.  
It would be interesting to seek other non-Markov systems with a c.i.i.d. structure.

\vspace{0.5cm}
{\bf Acknowledgements:} DB acknowledges support from CONACYT (Mexico) Grant CF2019/10872 and from the University of Paris-Saclay, France (scientific missionary). SNM acknowledges support from ANR Grant No. ANR-23-CE30-0020-01 EDIPS. The authors thank the Isaac Newton Institute (Cambridge, UK) for hospitality during a stay where this work was initiated.

\appendix

\section{Large $t$ asymptotics of $a_{\lambda=0}(t)$ in Eq. (\ref{ac2c4})}\label{sec:appA}

To compute $a_0(t)$ from the general exact expression in Eq. (\ref{ac2c4}), we first need to consider the quadratic coefficient given by Eq.~(\ref{var.1b}) and
use the definition of $E_1(z)= \int_z^{\infty} \frac{dx}{x}\, e^{-x}$. One gets for large $z=rt$,
\begin{equation}
\frac{c_2^2(z)}{(D/r)^2}= \left[\ln z + \gamma_E + \frac{e^{-z}}{z}+\ldots\right]^2
= \ln^2 z+ 2\gamma_E\, \ln z + \gamma_E^2 + O\left((\ln z)\, e^{-z}\right)\, .
\label{r2t.2}
\end{equation}
Regarding the term $c_4(z)$ given by Eq. (\ref{num.1b}), 
%that reads
%\begin{equation}
%r_1(t)= 2\, \left[-1+ \gamma_E + E_1(rt) + \ln (rt)+ e^{-rt}+ \frac{1}{2}\,
%_1F_1^{(2,0,0)}(0,1,-rt)\right]\ .
%\label{r1t.1}
%\end{equation}
it can be expanded at large $z$ as
\begin{equation}
\frac{c_4(z)}{(D/r)^2}=  \ln z + \gamma_E-1 + \frac{1}{2}\ _1F_1^{(2,0,0)}(0,1,-z)+ O\left(e^{-z}\right)\, .
\label{r1t.2}
\end{equation}
So, the question is how the term $_1F_1^{(2,0,0)}(0,1,-z)$ behaves for large $z$.
Unfortunately, Mathematica is unable to extract the asymptotic large $t$ behavior of this term.
So, we need to use the following trick. We note that by definition
\begin{equation}
_1F_1^{(2,0,0)}(0,1,-z)= \partial_b^2 M(b,1,-z)\Big|_{b=0}\, ,
\label{F_def}
\end{equation}
where $M(b,c,u)$ is defined in Eq. (\ref{Kummer.1}). To evaluate \eqref{F_def} for $z\gg1$,
we first use the asymptotic behaviour of $M(b,c,u)$ for large negative $u$ as given in 
Eq. (\ref{asymp.1}). This gives for large $t$
\begin{equation}
M(b, 1, -z) = \frac{1}{\Gamma(1-b)}\, z^{-b} + O\left(z^{-b-1}\right)\, .
\label{M1_asymp.1}
\end{equation}
We now expand the rhs of \eqref{M1_asymp.1} for small $b$ (but fixed large $z$) and from the
coefficient of $b^2$, we can read off the second derivative at $b=0$ in Eq. (\ref{F_def}).
This gives,
\begin{equation}
_1F_1^{(2,0,0)}(0,1,-z)= \ln^2z+ 2\gamma_E \ln z +\gamma_E^2-\frac{\pi^2}{6}
+ {\rm h.o.t}
\label{F1_leading.1}
\end{equation}
where the higher order terms (h.o.t) decreases with large $z$. Consequently, 
we find that Eq. \eqref{r1t.2} behaves as
\begin{equation}
\frac{c_4(z)}{(D/r)^2}= \frac{1}{2}\ln^2z + (1+\gamma_E)\ln z + \frac{1}{2}\gamma_E^2+\gamma_E -1-\frac{\pi^2}{12}+ {\rm h.o.t}
\label{r1t.3}
\end{equation}
where h.o.t again decreases for large $z$. Substituting \eqref{r1t.3} and \eqref{r2t.2}
in Eq. (\ref{ac2c4}) we get the final asymptotic tail of $a_0(t)$,
\begin{equation}
a_0(t) = \frac{1}{\ln z} - \frac{4+\gamma_E+ \frac{\pi^2}{12}}{\ln^2 z} +\ldots\, .
\label{AT_asymp_app}
\end{equation}


\begin{thebibliography}{99}

\bibitem{EM2011} M. R. Evans and S. N. Majumdar,
{\em Diffusion with stochastic resetting}, Phys. Rev. Lett. {\bf 106},
160601 (2011).

\bibitem{P2015}
A. Pal, {\em Diffusion in a potential landscape with stochastic resetting}, Phys. Rev. E {\bf 91}, 012113 (2015).

\bibitem{MSS2015}
S, N. Majumdar, S. Sabhapandit, and Gr\'egory Schehr,
{\em Dynamical transition in the temporal relaxation of stochastic processes under resetting}, Phys. Rev. E {\bf 91}, 052131 (2015).

\bibitem{EM2011b}
M. R. Evans and S. N. Majumdar,
{\em Diffusion with optimal resetting},
J. Phys. A: Math. Theor. {\bf 44} 435001 (2011).

\bibitem{NG2016}
A. Nagar and S. Gupta, {\em Diffusion with stochastic resetting at power-law times}, Phys. Rev. E {\bf 93}, 060102 (2016). 

\bibitem{EM2016}
S. Eule and J. J. Metzger, {\em Non-equilibrium steady states of stochastic processes with intermittent resetting}, New J.
Phys. {\bf 18}, 033006 (2016). 

\bibitem{PKR2019}
A. Pal, L. K\'usmierz, and S. Reuveni, {\em Time-dependent density of diffusion with stochastic resetting is invariant to return
speed}, Phys. Rev. E {\bf 100}, 040101 (2019). 

\bibitem{BS2020}
A. S. Bodrova and I. M. Sokolov, {\em Resetting processes with noninstantaneous return}, Physical Review E {\bf 101}, 052130 (2020).

\bibitem{EMS2020}
M. R. Evans, S. N. Majumdar, and G. Schehr, {\em Stochastic resetting and applications}, J. Phys. A {\bf 53}, 193001 (2020).

\bibitem{TPSRR2020}
O. Tal-Friedman, A. Pal, A. Sekhon, S. Reuveni, and Y. Roichman,
{\em Experimental Realization of Diffusion with Stochastic Resetting},  J. Phys. Chem. Lett. {\bf 11}, 7350 (2020).

\bibitem{BBPMC2020}
B. Besga, A. Bovon, A. Petrosyan, S. N. Majumdar, and S. Ciliberto,
{\em Optimal mean first-passage time for a Brownian searcher subjected to resetting: experimental and theoretical results},
Phys. Rev. Research {\bf 2}, 032029(R) (2020).

\bibitem{FBPCM2021}
F. Faisant, B. Besga, A. Petrosyan, S. Ciliberto, and
S. N. Majumdar,
{\em Optimal mean first-passage time of a Brownian searcher with resetting in one and two dimensions: experiments, theory and numerical tests}, J. Stat. Mech. 113203 (2021).

\bibitem{EM2014} M. R. Evans and S. N. Majumdar,
{\em Diffusion with resetting in arbitrary spatial dimension},
J. Phys. A {\bf 47}, 285001 (2014).

\bibitem{GMS2014} S. Gupta, S.~N. Majumdar, and G. Schehr,
{\em Fluctuating interfaces subject to stochastic
resetting}, Phys. Rev. Lett.,
{\bf 112}, 220601 (2014).


\bibitem{FE2017}
R. Falcao and M. R. Evans, {\em Interacting Brownian motion with resetting}, J. Stat. Mech. 023204 (2017). 

\bibitem{BKP2019}
U. Basu, A. Kundu, and A. Pal, {\em Symmetric exclusion process under stochastic resetting},
Phys. Rev. E {\bf 100}, 032136 (2019).


\bibitem{MMS2020}
M. Magoni, S. N. Majumdar, and G. Schehr, {\em Ising model with stochastic resetting},
Phys. Rev. Research,  {\bf 2}, 033182 (R) (2020).

\bibitem{MCG2024}
R. Majumder, R. Chattopadhyay, and S. Gupta,
{\em Kuramoto model subject to subsystem resetting: How resetting a part of the system may synchronize the whole of it}, 
Phys. Rev. E {\bf 109}, 064137 (2024).

\bibitem{BMP2025}
A. Biswas, S. N. Majumdar, and A. Pal, {\em Target search optimization by threshold resetting}, Phys. Rev. Lett. {\bf 135}, 227101‌ (2025).

\bibitem{NG2023}
A. Nagar and S. Gupta, 
{\em Stochastic resetting in interacting particle systems: a review}
J. Phys. A: Math. Theor.  {\bf 56}, 283001 (2023). 

\bibitem{BLMS2023} M. Biroli, H. Larralde, S. N. Majumdar, and G. Schehr,
{\em Extreme statistics and spacing distribution in a Brownian gas
correlated by resetting}, Phys. Rev. Lett., {\bf 130}, 207101 (2023).

%\bibitem{BMS2023} M. Biroli, S. N. Majumdar, G. Schehr,
%{\em Critical number of walkers for diffusive search processes with resetting},
%Phys. Rev. E, {\bf 107}, 064141 (2023).

\bibitem{BLMS2024} M. Biroli, H. Larralde,, S. N. Majumdar, and G. Schehr,
{\em Exact extreme, order and sum statistics in a class of strongly correlated system},
Phys. Rev. E {\bf 109}, 014101 (2024). 

\bibitem{BKMS2024} M. Biroli, M. Kulkarni, S. N. Majumdar, and G. Schehr,
{\em Dynamically emergent correlations between particles in a
switching harmonic trap},
Phys. Rev. E {\bf 109}, L032106 (2024).

\bibitem{SM2024} S. Sabhapandit and S. N. Majumdar,
{\em Noninteracting particles in a harmonic trap with a stochastically driven center},
J. Phys. A: Math. Theor. {\bf 57}, 335003 (2024).

\bibitem{MMS2025} N. Mesquita, S. N. Majumdar, and S. Sabhapandit,
{\em Dynamically emergent correlations in a Brownian gas with diffusing diffusivity},
J. Stat. Mech. 103207‌ (2025).


\bibitem{DBMS2025} G. de Mauro, M. Biroli, S. N. Majumdar, and G. Schehr,
{\em Dynamically emergent correlations in Brownian particles subject to simultaneous non-Poissonian resetting protocols},
Phys. Rev. E‌ {\bf 113}, 014120‌ (2026).

\bibitem{BCKMPS2025} M. Biroli, S. Ciliberto, M. Kulkarni, S. N. Majumdar, A. Petrosyan, and G. Schehr,
{\em Experimental evidence for strong emergent correlations between particles in a switching trap},
arXiv: 2508.07199 (2025).

\bibitem{MCPL2022}
M. Magoni, F. Carrolo, G. Perfetto, and I. Lesanovsky, {\em 
Emergent quantum correlations and collective behavior in 
non-interacting quantum systems subject to stochastic resetting}, Phys. Rev. A {\bf 106}, 052210 (2022).

\bibitem{SLP2025}
D Soldner, I Lesanovsky, and G Perfetto, {\em 
Nonanaliticities and ergodicity breaking in noninteracting many-body dynamics via stochastic resetting and global measurements}, arXiv: 2510.11450 (2025).

\bibitem{KMS2024} M. Kulkarni, S. N. Majumdar, and S. Sabhapandit,
{\em Dynamically emergent correlations in bosons via quantum resetting}, J. Phys. A {\bf 58}, 105003 (2025).

\bibitem{MKMS2025}  N. Mesquita, M. Kulkarni, S. N. Majumdar, and S. Sabhapandit,
{\em Dynamically generated correlations in a trapped bosonic gas via frequency quenches},
J. Phys.: Conf. Ser.‌ {\bf 3152}, 012011‌ (2025).

\bibitem{BS2014} D. Boyer and C. Solis-Salas, {\em Random walks with preferential relocations to places visited in the past and
their application to biology}, Phys. Rev. Lett. {\bf 112}, 240601 (2014).

\bibitem{BEM2017} D. Boyer, M. R. Evans, and S. N. Majumdar, {\em Long time scaling behaviour for
diffusion with resetting and memory}, J. Stat. Mech. 023208 (2017).

\bibitem{MU2019} C. Mailler and G. Uribe Bravo,
{\em Random walks with preferential relocations and fading memory: a study through random recursive trees},
J. Stat. Mech. 093206 (2019).

\bibitem{GM2005}
A. O. Gautestad and I. Mysterud, {\em Intrinsic scaling complexity in animal dispersion and abundance}, Am. Nat. {\bf 165}, 44 (2005).

\bibitem{SKWB2010}
C. Song, T. Koren, P. Wang, and A.-L. Barab\'asi, {\em Modelling the scaling properties of human mobility}, Nature Physics {\bf 6}, 818 (2010).

\bibitem{F2013}
W. F. Fagan, M. A. Lewis, M. Auger-M\'eth\'e, T. Avgar, S. Benhamou, G. Breed, L. LaDage, U. E. Schl${\rm \ddot{a}}$gel, W.-w. Tang,
Y. P. Papastamatiou, et al., {\em Spatial memory and animal movement}, Ecol. Lett. {\bf 16}, 1316 (2013).

\bibitem{MFM2014}
J. Merkle, D. Fortin, and J. M. Morales, {\em A memory-based foraging tactic reveals an adaptive mechanism for restricted
space use}, Ecol. Lett. {\bf 17}, 924 (2014).

\bibitem{FBMFM2021}
A. Falc\'on-Cort\'es, D. Boyer, E. Merrill, J. L. Frair, and J. M. Morales, {\em Hierarchical, memory-based movement models for translocated {E}lk ({C}ervus canadensis)}, Front. Ecol. Evol {\bf 9}, 702925 (2021).

\bibitem{R2021}
N. Ranc, P. R. Moorcroft, F. Ossi, and F. Cagnacci, {\em Experimental evidence of memory-based foraging decisions in a large
wild mammal}, Proc. Natl. Acad. Sci. USA {\bf 118}, e2014856118 (2021).

\bibitem{RCM2022}
N. Ranc, F. Cagnacci, and P. R. Moorcroft, {\em Memory drives the formation of animal home ranges: evidence from a
reintroduction}, Ecol. Lett. {\bf 25}, 716 (2022).

\bibitem{AS1965} M. Abramowitz and I. A. Stegun (Eds.), {\em Handbook of Mathematical Functions with Formulas, Graphs, and Mathematical 
Tables} (Dover, New York, 1965).

\bibitem{B1996}
G. I. Barenblatt, {\em Scaling, Self-similarity, and Intermediate Asymptotics} (Cambridge, Cambridge, UK, 1996).

\end{thebibliography}
\end{document}